\documentclass[journal]{IEEEtran}

\ifCLASSINFOpdf
\else
\fi

\usepackage{multirow}
\usepackage[table,xcdraw]{xcolor}
\usepackage{amsmath}
\usepackage{booktabs}
\usepackage{cite}
\usepackage{amssymb}
\usepackage{textcomp}
\usepackage{hyperref}
\usepackage{graphicx}
\usepackage{amssymb}
\usepackage{cleveref}
\usepackage{CJKutf8}
\usepackage{bbding}
\crefname{figure}{fig}{figures}
\Crefname{figure}{Fig}{Figures}
\begin{document}
\begin{CJK}{UTF8}{gbsn}
\makeatletter
\newcommand{\rmnum}[1]{\romannumeral #1}
\newcommand{\Rmnum}[1]{\expandafter\@slowromancap\romannumeral #1@}
\makeatother

%
\title{Implicit Neural Representation-Based MRI Reconstruction Method with sensitivity map Constraints}
%
%
%

\author{~\IEEEmembership{Lixuan~Rao}, ~\IEEEmembership{Xinlin~Zhang*,~Member,~IEEE}, ~\IEEEmembership{Yiman~Huang}, ~\IEEEmembership{Tao~Tan}, ~\IEEEmembership{Tong~Tong*}

\thanks{This work was supported by supported in part by the National Natural Science Foundation of China under Grant 62401151 and 62171133, the Natural Science Foundation of Fujian Province under the Grant 2024108138, and the Fuzhou University research start-up funding. *Corresponding author: Xinlin Zhang and Tong Tong.}
\thanks{Lixuan~Rao, Xinlin~Zhang*, and Tong~Tong* are with College of Physics and Information Engineering and Fujian Key Lab of Medical Instrumentation Pharmaceutical Technology, Fuzhou University, Fuzhou 350108, China.}
\thanks{Yiman~Huang is with College of Electrical Engineering, Zhejiang University, Hangzhou 310058, China.}
\thanks{Tao~Tan is with Faulty of Applied Science, Macao Polytechnic University. 999078, Macao Special Administrative Region of China, China.}}

\maketitle 

\begin{abstract}
Magnetic Resonance Imaging (MRI) is a widely utilized diagnostic tool in clinical settings, but its application is limited by the relatively long acquisition time. As a result, fast MRI reconstruction has become a significant area of research. In recent years, Implicit Neural Representation (INR), as a scan-specific method, has demonstrated outstanding performance in fast MRI reconstruction without fully-sampled images for training. High acceleration reconstruction poses a challenging problem, and a key component in achieving high-quality reconstruction with much few data is the accurate estimation of coil sensitivity maps. However, most INR-based methods apply regularization constraints solely to the generated images, while overlooking the characteristics of the coil sensitivity maps. To handle this, this work proposes a joint coil sensitivity map and image estimation network, termed INR-CRISTAL. The proposed INR-CRISTAL introduces an extra sensitivity map regularization in the INR networks to make use of the smooth characteristics of the sensitivity maps. Experimental results show that INR-CRISTAL provides more accurate coil sensitivity estimates with fewer artifacts, and delivers superior reconstruction performance in terms of artifact removal and structure preservation. Moreover, INR-CRISTAL demonstrates stronger robustness to automatic calibration signals and the acceleration rate compared to existing methods.
\end{abstract}


%
\IEEEpeerreviewmaketitle

\section{Introduction}

\IEEEPARstart{M}{agnetic} Resonance Imaging (MRI), a widely used diagnostic tool in clinical medicine \cite{MRItechnology}, often requires lengthy scanning procedures, which can cause patient discomfort. Moreover, prolonged scans increase the risk of image distortion due to natural body movements such as heartbeat and respiration. Consequently, accelerating MRI through undersampling and reconstruction has become a key research focus in the field \cite{fastMRI}, \cite{liang2007spatiotemporal}.

Parallel imaging (PI) \cite{PI}, \cite{SENSE}, \cite{GRAPPA} has been widely used in commercial MRI scanners for acceleration. The PI reconstruction methods can be broadly categorized into image domain \cite{SENSE}, \cite{JSENSE}, \cite{PFISTA}, and k-space reconstruction approaches \cite{GRAPPA}, \cite{SPIRiT}, \cite{PILS}. The former relies on pre-calculated coil sensitivity maps \cite{ESPRIRT}, \cite{JDSI}, and the latter learns kernels for reconstruction from the auto-calibration signal (ACS) \cite{AUTO-SMASH}. Compressed sensing further accelerates MRI acquisition by non-uniform sampling and imposing appropriate constraints on the reconstruction models, such as image sparsity \cite{CS_1}, \cite{CS_2}, low rankness \cite{LORAKS}, \cite{ACLORAKS}, \cite{PLORAKS}, \cite{SAKE}, \cite{calibrationless_SAKE}, or a combination of both \cite{Xinlin1}, \cite{Xinlin2}.

With the advancement of deep learning, fast imaging techniques based on these methods have demonstrated considerable promise, offering improvements in both reconstruction speed and image quality compared to traditional optimization approaches \cite{DLMRI_1}, \cite{DLMRI_2}, \cite{DLMRI_3}, \cite{DLMRI_4}, \cite{PFISTA-NET}. Nevertheless, most deep learning models necessitate large volumes of fully sampled prior data for pre-training, leading to extended training times \cite{DLMRI_5}, \cite{DLMRI_6}, \cite{DLMRI_7}, \cite{MoDL}, and are also challenged by issues of data generalization when there is a mismatch between the training and inference data distributions.

In recent years, implicit neural representation (INR) has garnered significant attention as a self-supervised approach \cite{NERF}, \cite{NERP}, \cite{NERFSF}, \cite{NEX}, \cite{INR1}, \cite{NIK}, \cite{IMJENSE}, \cite{CINEJENSE}. Unlike traditional discrete representations that store signal values or features explicitly, INR defines a continuous generator function that maps input coordinates directly to the corresponding values in the output space. Typically, this function is parameterized by a fully connected neural network, most commonly multi-layer perceptrons (MLPs), enabling the representation of complex signals such as images, videos, and even 3D scenes with high fidelity and resolution. Owing to their resolution independence, memory efficiency, and compatibility with gradient-based optimization, INR-based models have shown remarkable promise in a variety of tasks, including image reconstruction, novel view synthesis, and medical image modeling.

In the field of MRI reconstruction, INR has been applied in various imaging scenarios, yielding promising results \cite{NERP}, \cite{IMJENSE},  \cite{CINEJENSE}, \cite{INR_SURVEY}, \cite{INFusion}, \cite{INR_2025}, \cite{chu2025highly}. Implicit neural representation learning with prior embedding (NeRP) \cite{NERP} exploited INR to map image coordinates to corresponding intensity values for MRI reconstruction. However, NeRP requires a fully sampled image prior of the same subject for pre-training. To address this issue, Feng et al. proposed a self-supervised implicit representation for joint coil sensitivity and image estimation (IMJENSE) \cite{IMJENSE}, which jointly estimates coil sensitivities and MRI images. In IMJENSE, the image was generated by MLPs, and the sensitivity map was parameterized by polynomials. Expanding on IMJENSE, CineJENSE \cite{CINEJENSE} replaced the polynomials for coil sensitivity estimation with another INR network and extended to cardiac MRI reconstruction. In addition, INR and its varieties have been applied to different MRI reconstruction problems such as dynamic MRI, mapping, etc \cite{INR_SURVEY}, \cite{INFusion}, \cite{INR_2025}, \cite{chu2025highly}.

However, few INR-based methods consider leveraging the smoothness of the coil sensitivity maps and imposing constraints on them during the training process. Many studies have shown that accurate sensitivity estimation is essential for high-quality MRI reconstruction using image domain PI methods \cite{JSENSE}, \cite{ESPRIRT}, \cite{JDSI}. It has been shown that incorporating appropriate constraints on sensitivity maps can facilitate its accurate estimation and further reduce the error of reconstructed images, in optimization-based reconstruction \cite{JSENSE}, \cite{iSENSE}, and supervised deep learning reconstruction \cite{jense-pro}. Although the polynomial sensitivity map modeling of IMJENSE can be regarded as an implicit smooth constraint, the sensitivity maps generated by IMJENSE have expanded support to the actual image support, with little high-frequency detail information, resulting in suboptimal results (Fig.~\ref{fig:show_csm}~(b)). Coil sensitivity maps could be successfully modeled by MLPs in CineJENSE (which is reduced to two dimensions), providing sensitivity predictions closer to the ground truth. However, when the ACS data is collected in minimal quantities, CineJENSE fails to produce accurate sensitivity maps with severe artifacts (red arrows in Fig.~\ref{fig:show_csm}~(c)), finally leading to degraded image quality.

Therefore, an INR-based method with coil sensitivity regularization image and sensitivity estimation algorithm (INR-CRISTAL) is proposed for PI reconstruction in this work. Two MLPs estimated the composite MRI image and corresponding coil sensitivity maps, respectively. The proposed method leverages the smooth characteristics of the sensitivity map by introducing an extra sensitivity map regularization in the INR networks. This regularization term on sensitivity can be flexibly modeled according to different applications. We discussed the effects of different regularization terms on reconstruction results and provided recommended regularization to obtain promising reconstructions. Experimental results demonstrate that the proposed INR-CRISTAL outperforms the state-of-the-art optimization-based or scan-specific reconstruction methods regarding aliasing artifact suppression. Moreover, the proposed INR-CRISTAL exhibits robustness to the ACS and reduction factor (R). The INR-CRISTAL still achieves the lowest reconstruction error in the case that the ACS region is too limited or R is relatively high.

\section{Related Works}

\subsection{Coil Sensitivity Map Optimization }
Reliable coil sensitivity maps are essential for high-quality MRI reconstruction in PI. However, with limited ACS lines, conventional methods like ESPIRiT often fail to produce accurate sensitivities, leading to severe artifacts and degraded image quality \cite{JDSI}, \cite{calibrationless_SAKE}.

To overcome this limitation, joint estimation of images and sensitivities was introduced in methods such as JSENSE \cite{JSENSE}, using polynomial parameterizations. More recent deep learning approaches like \cite{JDSI} embed sensitivity estimation into iterative reconstruction. These strategies effectively suppress artifacts and improve reconstruction quality, especially under limited calibration conditions.
\begin{figure}
    \centering 
    \includegraphics[width=0.48\textwidth]{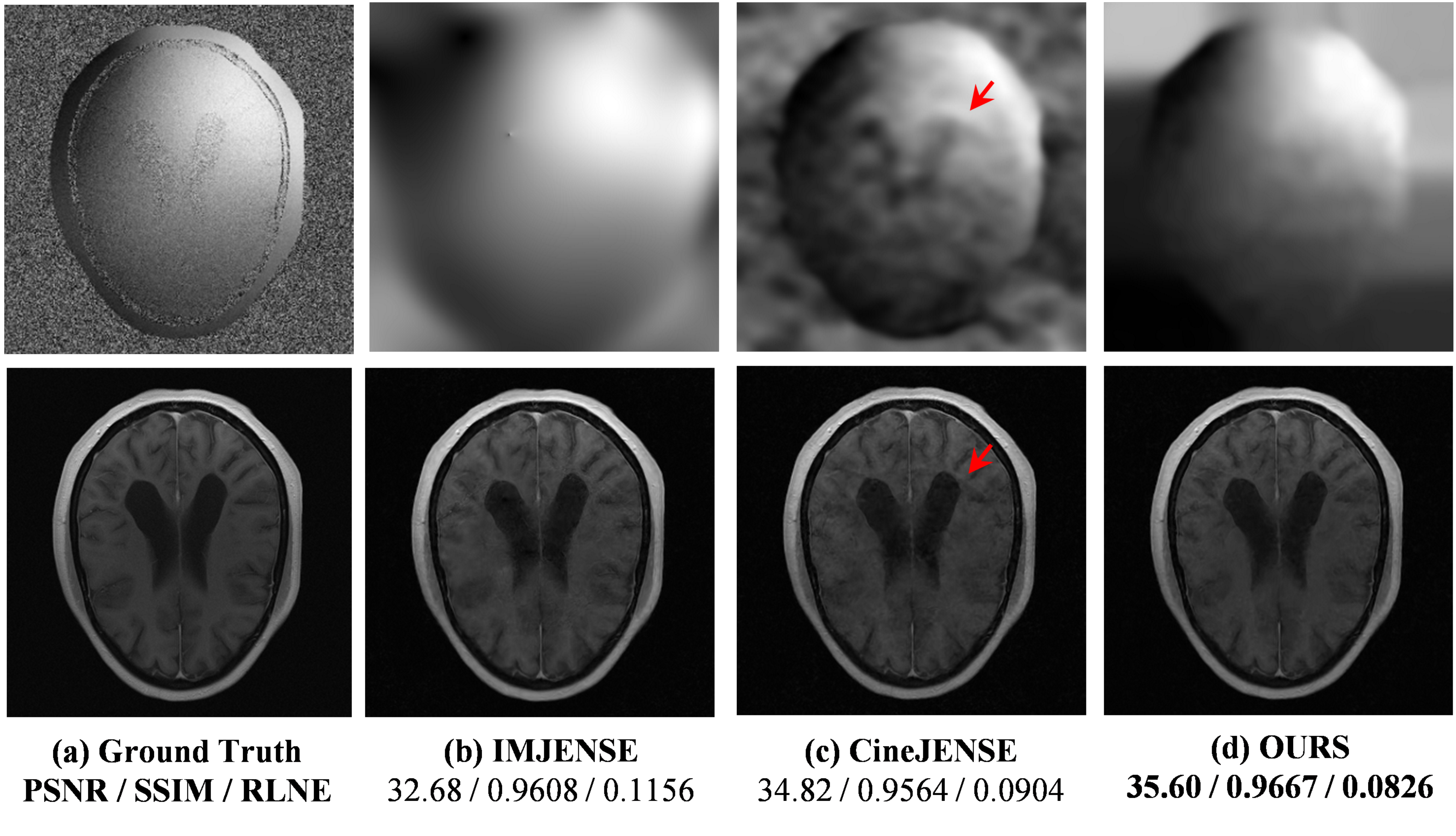}
    \caption{Different methods for coil sensitivity maps and reconstructed images constructed from brain data uniform sampling (ACS=8, R=5). (a) Ground truth includes both sensitivity maps and image; the ground truth sensitivity maps were generated by dividing individual fully sampled coil images by the SOS image of fully sampled multi-coil images. (b)-(d) Sensitivity maps and images estimated using different methods, respectively. Note that the CineJENSE used here is a 2D version implemented by us and can be considered as the proposed INR-CRISTAL without sensitivity regularization. }
    \label{fig:show_csm}
\end{figure}

The analysis revealed the shortcomings of various methods, as illustrated in Fig.~\ref{fig:show_csm}. Specifically, the sensitivity maps estimated through polynomial joint estimation \cite{JSENSE} and ESPIRiT \cite{ESPRIRT} lack high-frequency details and frequently mismatch the actual sensitivity values, with the estimated sensitivity map often exceeding the proper distribution range. Meanwhile, the coil sensitivity map generated by INR is prone to artifacts; both of these factors contribute to image quality degradation.

In summary, relying solely on the aforementioned methods is insufficient to reconstruct sensitivity maps that not only avoid artifacts but also closely match the actual sensitivity range and preserve the details of the sensitivity map.

\subsection{Implicit Neural Representation}
The reconstruction models of IMJENSE can be summarized as the optimization problem shown below:
\begin{equation}
    \underset{\alpha, \beta}{\operatorname{argmin}} \frac{1}{2} \sum_{j=1}^{N}\left\|\mathbf{Y}_{j}-\boldsymbol{\mathcal{U} \mathcal { F }} \mathbf{S}_{\beta j} \odot \mathbf{X}_{\alpha}\right\|_{2}^{2}+\lambda \mathcal{R}(\mathbf{X_{\alpha}}),
    \label{eq:1}
\end{equation}
where $\mathbf{X_{\alpha}} $ denotes the image to be reconstructed, $\mathbf{S}_{\beta j}$
is the sensitivity map of the $j^{\text{th}}$ channel, and  $\mathbf{S_{\beta}}=[\mathbf{S}_{\beta 1}, \ldots,\mathbf{S}_{\beta j}, \ldots, \mathbf{S}_{\beta N}]$.  Then, symbol ${\alpha}$ represents the parameters to be optimized in the implicit neural network for the image, while ${\beta}$ represents the parameters to be optimized in the model for the sensitivity maps. $\mathbf{Y}_{j}$ denotes the acquired k-space data with zero-filling of $j^{\text{th}}$ coil. The symbol $\mathcal{F} $ denotes the operator that performs the 2D Fourier transform on the multi-channel image; $\mathcal{U}$, the operator that performs undersampling and fills the unsampled positions with zeros, $\mathcal{R}$  the regularization operator, and $\odot$ represents the Hadamard product. The parameter $\lambda$ controls the strength of the regularization constraint, and $\|\cdot\|_{2}^{2}$ is the square of the l2-norm of a vector.

From Eq. \eqref{eq:1}, it can be seen that the optimization problem of IMJENSE does not impose any constraints on the sensitivity map. 

\begin{figure*}
    \centering 
    \includegraphics[width=\textwidth]{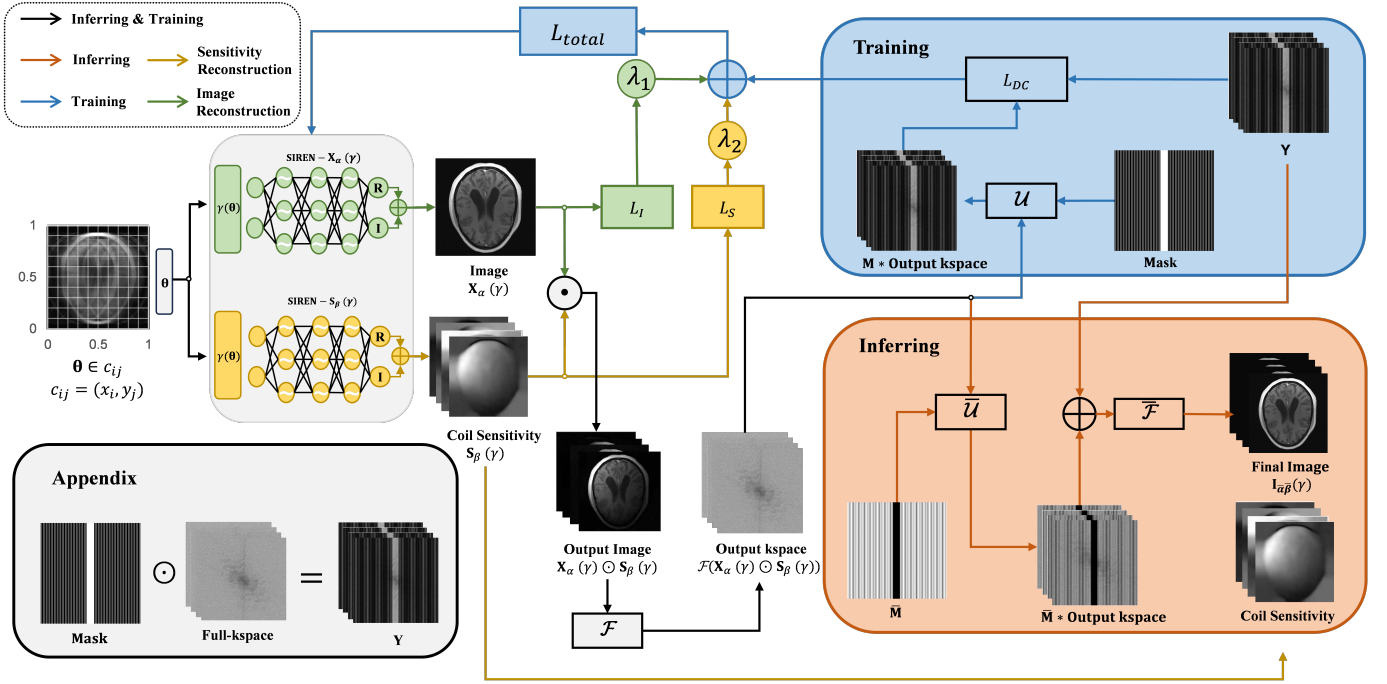}
    \caption{Flowchart of the proposed method. First, the coordinates are determined by the positions in the image. The $\boldsymbol{\theta}$ is a matrix that represents the coordinate set. After Fourier feature embedding, it is used as the input to two networks. The INR then generates a single-channel image and a sensitivity map. During the training process, the loss of consistency of the data $L_{DC}$, the loss of total variation of the image $L_{I}$, and the loss of total variation of the sensitivity $L_{S}$ are used to simultaneously optimize the two networks. During inference, a combined approach is used to enhance the accuracy of the k-space.}
    \label{fig:progress_picture}
\end{figure*}

\section{Method}

\subsection{Problem Formulation }
In INR-based parallel imaging methods, different strategies have been employed to generate coil sensitivity maps. For instance, IMJENSE estimates coil sensitivity maps using continuous polynomial functions, while CineJENSE leverages INR to reconstruct them. Although both methods impose implicit constraints on the generated sensitivity maps, they often suffer from inaccuracies when the number of ACS lines is limited, which subsequently degrades the final image quality.

A key limitation shared by both approaches is the absence of explicit constraints on the sensitivity maps. Many studies have shown that incorporating appropriate regularization on coil sensitivity maps can significantly improve reconstruction performance, \cite{iSENSE}, \cite{jense-pro}. Therefore, we introduce a coil sensitivity regularization to Eq. \eqref{eq:1}:
\begin{equation}
    \underset{\alpha, \beta}{\operatorname{argmin}} \frac{1}{2} \sum_{j=1}^{N}\left\|\mathbf{Y}_{j}-\mathcal{U} \mathcal{F} \mathbf{S}_{\beta j} \odot \mathbf{X}_{\alpha}\right\|_{2}^{2}+\lambda_{1} \mathcal{R}(\mathbf{X}_{\alpha})+\lambda_{2} \mathcal{R}(\mathbf{S_{\beta}}),
    \label{eq:2}
\end{equation}
where $\lambda_{1}$ and $\lambda_{2}$ separately control the regularization strength for the image and sensitivity maps.

\subsection{Coil Sensitivity Constraints}
\label{Coils Sensitivity constraint}
Various regularization strategies have been proposed for coil sensitivity maps. Since the coil sensitivity maps appear smooth in the image domain, some methods are based on the assumption that the coil sensitivities are sparse in the Chebyshev polynomial basis or in the Fourier domain \cite{iSENSE}, \cite{jense-pro}. The regularization term $\mathcal{R}(\mathbf{S}_{\beta})$ can be formulated as:
\begin{equation}
\mathcal{R}_{L1}(\mathbf{S_{\beta}}) = \sum_{j=1}^{N}\left \| \mathcal{F}(\mathbf{S}_{\beta j}) \right \| _{1},
    \label{eq:l1}
\end{equation}
where $\|\cdot\|_{1}$ denotes the l1-norm of a vector.

Another approach models the sensitivity maps as rank-deficient matrices and imposes the low-rankness constraint by minimizing the nuclear norm of the coil sensitivity matrix \cite{iSENSE}, \cite{jense-pro}. This regularization is expressed as:
\begin{equation}
\mathcal{R}_{LR}(\mathbf{S_{\beta}}) = \sum_{j=1}^{N}\left \| \mathbf{S}_{\beta j} \right \| _{*},
    \label{eq:lowrank}
\end{equation}
where $\|\cdot\|_{*}$ denotes the nuclear norm.

In addition, total variation is also an effective way to constrain image smoothness, since it is the sum of the absolute variations or gradients in the image. Therefore, we propose to impose a total variation (TV) penalty on the sensitivity maps. It is trivial yet very effective, and to the best of our knowledge, this is the first application of the TV constraint on sensitivity maps. This is formulated as:
\begin{equation}
\mathcal{R}_{TV}(\mathbf{S_{\beta}}) =  \sum_{j=1}^{N} \left\|\mathcal{G}\left(\mathbf{S}_{\beta j}\right)\right\|_{1},
    \label{eq:tv}
\end{equation}
where $\mathcal{G}$ denotes the spatial gradient operator. 

In practical reconstruction, appropriate regularization strategies can be selected for the coil sensitivity maps according to specific requirements, thereby improving the quality of the reconstructed images. A comprehensive discussion of different regularization strategies is provided in \ref{sec:reconstruction_constraints}.

\subsection{INR-CRISTAL }
The Fig.~\ref{fig:progress_picture} illustrates the proposed method, where the coordinates, represented by matrix $\boldsymbol{\theta}$, are determined based on their positions in the image and used as input to two networks after Fourier feature embedding. The INR generates both a composited image and a sensitivity map, with the training process optimizing the two networks simultaneously using the data consistency loss $L_{DC}$, the image total variation loss $L_{I}$, and sensitivity total variation loss  $L_{S}$. During inference, a combined approach is employed to enhance the accuracy of the k-space.

\subsubsection{Network Construction}
To address the inherent low-frequency bias of MLP-based INR \cite{rahaman2019spectral}, which often hinders the reconstruction of high-frequency details, we employ the sine representation network (SIREN \cite{SIREN}). By introducing periodic sine activations and adjusting the spatial frequency parameter $w_0$, SIREN significantly enhances the network’s capacity to model high-frequency components, thereby improving the fidelity of image reconstruction in fine structural regions.

In our framework, two separate networks, $\mathbf{X}_{\alpha}$ and $\mathbf{S}_{\beta}$, are constructed to represent the composited image and the coil sensitivity maps, respectively. For the $j^{\text{th}}$ coil, the estimated sensitivity map is denoted as $\mathbf{S}_{\beta j}$, with the spatial coordinates $(x, y)$ serving as the network input. Equation~\eqref{eq:2} is thus extended accordingly:
\begin{equation}
\begin{aligned}
&\underset{\alpha, \beta}{\operatorname{argmin}} \frac{1}{2} \sum_{j=1}^{N}\left\|\mathbf{Y}_{j}-\mathcal{U} \mathcal{F} \mathbf{S}_{\beta j}(\boldsymbol{\theta}) \odot \mathbf{X}_{\alpha}(\boldsymbol{\theta})\right\|_{2}^{2}
  \\  
&+ \lambda_{1}\left\|\mathcal{G}\left(\mathbf{X}_{\theta}(\boldsymbol{\gamma})\right)\right\|_{1}+{\lambda_{2}} \mathcal{R}(\mathbf{S_{\beta}(\boldsymbol{\theta})}),
\end{aligned}
    \label{eq:3}
\end{equation}
where  $\boldsymbol{\theta}$ denotes the set of normalized spatial coordinates, and each $c_{ij} \in \boldsymbol{\theta}$ corresponds to the coordinate at the $i^{\text{th}}$ row and $j^{\text{th}}$ column, $c_{ij}=\left({x}_{i}, 
y_{j}\right),  x_{i} \in[0,1) \quad y_{j} \in[0,1)$. 
This problem has been transformed into optimizing parameters in the INR network by searching for solutions in the function space it represents, rather than directly manipulating the desired image. Since the INR network learns continuous functions, it implicitly incorporates continuity priors into the image representation, effectively suppressing noise and artifacts.

\subsubsection{Fourier Feature Embedding}
To enable the network to better learn high-frequency functions in low-dimensional problems, we apply Fourier feature embedding \cite{FFM} to the input coordinates 
$\boldsymbol{\theta}$ and then use the encoded coordinates as the input to the network. The specific method for establishing the Fourier feature embedding is as follows:
\begin{equation}
    \boldsymbol{\gamma}(\boldsymbol{\theta})=[\cos (2 \pi \mathbf{B}    \boldsymbol{\theta}), \sin (2 \pi \mathbf{B} \boldsymbol{\theta})]^{T},
    \label{eq:4}
\end{equation}
where the matrix $\mathbf{B}$ represents the coefficients of the Fourier feature transformation, and is a real-valued matrix of size $(\mathbf{E},\mathbf{N}_{\theta})$. The $\mathbf{E}$ represents the output size of the Fourier feature embedding, and $\mathbf{N}_{\theta}$ represents the dimensionality of the coordinates. Each entry in the matrix $\mathbf{B}$ is sampled from a Gaussian distribution $\mathcal{N}\left(0, \boldsymbol{\sigma}^{2}\right)$. Therefore Eq. \eqref{eq:3} can be extended as follows:
\begin{equation}
\begin{aligned}
    & \underset{\alpha, \beta}{\operatorname{argmin}} \frac{1}{2} \sum_{j=1}^{N}\left\|\mathbf{Y}_{j}-\mathcal{U} \mathcal{F} \mathbf{S}_{\beta j}(\boldsymbol{\gamma}) \odot \mathbf{X}_{\alpha}(\boldsymbol{\gamma})\right\|_{2}^{2} \\
    &+\lambda_{1}\left\|\mathcal{G}\left(\mathbf{X}_{\alpha}(\boldsymbol{\gamma})\right)\right\|_{1}+ {\lambda_{2}} \mathcal{R}(\mathbf{S_{\beta}(\boldsymbol{\gamma})}).
\end{aligned}
\label{eq:5}
\end{equation}

\subsubsection{Training and inferring}
The training of an implicit neural network-based magnetic resonance image reconstruction model with sensitivity map constraints is achieved by minimizing the total loss function $L_{total}$ to estimate the optimal parameters $\bar{\alpha} $ and $\bar{\beta}$ in the implicit neural network. The total loss function is defined as:
\begin{equation}
    \mathit{L}_{\mathit{total}}=\mathit{L}_{\mathit{DC}}+\lambda_{1} \times \mathit{L}_{\mathit{I}}+\lambda_{2} \times \mathit{L}_{\mathit{S}},
    \label{eq:6}
\end{equation}
so the total loss function can be divided into three parts: the data consistency loss $L_{DC}$, the image total variation loss $L_{I}$ \cite{TVLOSS}, and coil sensitivity constraint loss  $L_{S}$:
\begin{equation}
\mathit{L}_{\mathit{DC }}=\sum_{j=1}^{N}\left\|\mathbf{Y}_{j}-\mathcal{U} \mathcal{F} \mathbf{S}_{\beta j}(\boldsymbol{\gamma}) \odot \mathbf{X}_{\alpha}(\boldsymbol{\gamma})\right\|_{1},
    \label{eq:7}
\end{equation}
\begin{equation}
\mathit{L}_{\mathit{I }}=\left\|\mathcal{G}\left(\mathbf{X}_{\alpha}(\boldsymbol{\gamma})\right)\right\|_{1},
    \label{eq:8}
\end{equation}
\begin{equation}
\mathit{L}_{\mathit{S }}=\mathcal{R}(\mathbf{S_{\beta}(\boldsymbol{\gamma})}),
    \label{eq:9}
\end{equation}
where $\mathit{L}_{\mathit{S }}$ can choose different constraints according to different situations.

The loss function is minimized using the gradient descent algorithm and standard backpropagation, with the model consisting of two implicit neural networks, where the inputs are the coordinates encoded through Fourier feature embedding $\boldsymbol{\gamma}(\boldsymbol{\theta})$.

After training, the network learns the reconstruction representation of MRI images, allowing direct generation of the target image. To obtain more accurate k-space results, we combine the model's output with the sampled k-space data, optimizing the k-space, which is then transformed into the final image $\mathbf{I}_{\bar{\alpha} \bar{\beta}}$ through a 2D inverse Fourier transform:
\begin{equation}
\mathbf{I}_{\bar{\alpha}, \bar{\beta}}(\boldsymbol{\gamma})=\sum_{j=1}^{N}\left(\overline{\mathcal{F}}\left(\overline{\mathcal{U}} \mathcal{F}\left(\mathbf{S}_{\beta j}(\boldsymbol{\gamma}) \odot \mathbf{X}_{\bar{\alpha}}(\boldsymbol{\gamma})\right)\right)+\mathbf{Y}_{j}\right),
    \label{eq:11}
\end{equation}
where $\mathbf{X}_{\bar{\alpha}}$ and $\mathbf{S}_{\bar{\beta j}}$ represent the single-channel image and the sensitivity map of the $j^{\text{th}}$ channel, respectively, when the model is optimized; matrix $\mathbf{I}_{\bar{\alpha} \bar{\beta}}$ represents the final multi-channel composite image; symbol $\overline{\mathcal{F}}$ is the operator that performs a 2D fast inverse Fourier transform on the multi-channel image and, symbol $\overline{\mathcal{U}}$ is the operator that performs complementary undersampling with $\mathcal{U}$ (It means performing zero-filling at the positions where $\mathcal{U}$ samples $\overline{\mathcal{U}}$, and performing zero-filling at the positions where $\overline{\mathcal{U}}$ samples $\mathcal{U}$).

\section{Experimental Results}
\subsection{Datasets}
Four datasets are used in our experiments, including the brain and knee datasets. The detailed information is listed as follows:
\begin{enumerate}
    \item Dataset \Rmnum{1}: A 14-channel T1 POST human brain dataset was acquired using a 2.9T SIEMENS scanner with an original in-plane resolution of 640 × 320, later center-cropped to 320 × 320. The acquisition parameters were as follows: matrix size = 640 × 320, TR/TE = 250/2.64 ms, FOV = 440 × 220 mm², and slice thickness = 5 mm.
    \item Dataset \Rmnum{2}: A 20-channel T2 FLAIR human brain dataset was acquired using a 2.9T SIEMENS scanner with an original in-plane resolution of 640 × 320, later center-cropped to 320 × 320. The acquisition parameters were as follows: matrix size = 640 × 320, TR/TE = 9000/81 ms, FOV = 440 × 220 mm², and slice thickness = 7.5 mm.
    \item Dataset \Rmnum{3}: This dataset was obtained by scanning healthy subjects using a 32-channel head coil with a 3T SIEMENS scanner (SIEMENS Healthcare, Erlangen, Germany), which was then compressed into 4 channels \cite{coilcompress}. The specific scanning parameters are as follows: matrix size = 256 × 256, TR/TE = 6100/99 ms, FOV = 220 × 220 mm², slice thickness = 3 mm.
    \item Dataset \Rmnum{4}: A fully sampled 15-channel knee MRI dataset with a resolution of 368 × 640, acquired using a 2D turbo spin-echo sequence without fat suppression.
\end{enumerate}

Dataset \Rmnum{1}, \Rmnum{2}, and \Rmnum{4} are all obtained from the NYU fastMRI Initiative database \cite{fastmri_1}, \cite{fastmri_2}. The number of layers used in experiments for each dataset is specified in the corresponding sections. The reconstructed multi-channel images are combined and displayed using the sum-of-squares (SOS) method.

\subsection{Implementation Details}
 The SIREN network consists of 1 input layer, 6 hidden layers, and 1 output layer.  Each hidden layer has 256 neurons, and all layers except the output layer use periodic sine activation functions. The input layer does not use the aforementioned uniform distribution and the output layer generates the corresponding image intensity for a given coordinate. We trained the two SIREN networks using two Adam optimizers \cite{ADAM}, with an initial learning rate of $10^{-4}$  and a total of 1000 iterations. The proposed method is implemented on a server with Python 3.8.17, which is equipped with 8 NVIDIA GeForce RTX 4090 GPUs, each with 24 GB of memory.

\begin{figure}
    \centering 
    \includegraphics[width=0.48\textwidth]{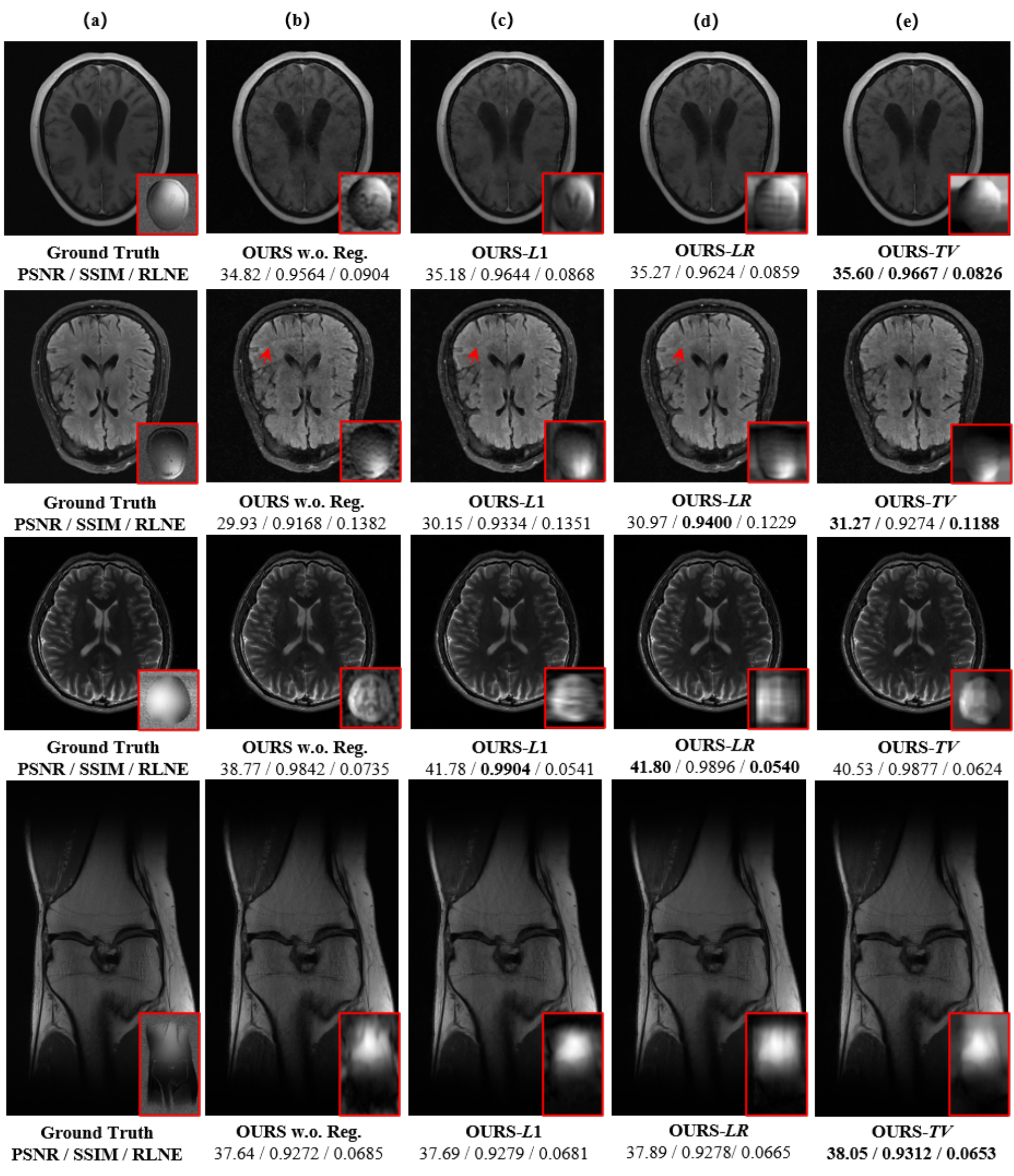}
    \caption{Different methods for coil sensitivity maps and reconstructed images from Dataset \Rmnum{1}, \Rmnum{2}, \Rmnum{3}, and \Rmnum{4} with uniform sampling ACS = 8 and acceleration factor R = 5 for Datasets \Rmnum{1}, \Rmnum{2}, \Rmnum{3}, and ACS = 24 and R = 5 for Dataset \Rmnum{4}.
    (a) Ground truth. (b) OURS w.o. Reg. (INR-CRISTAL without regularization). (c)-(e) Sensitivity maps and reconstructed images estimated using three different regularizations: $\mathcal{R}_{L1}$, $\mathcal{R}_{LR}$, and $\mathcal{R}_{TV}$, respectively. The images in the red box represent their respective sensitivity maps.}
    \label{fig:l1_lowrank}
\end{figure}

\subsection{Evaluation Metrics and Comparison Methods}
Quantitative metrics, including Peak Signal-to-Noise Ratio (PSNR), Structural Similarity Index (SSIM) \cite{SSIM}, and Relative L2 Norm Error (RLNE) \cite{RLNE}, are used in our experiments. Higher PSNR and SSIM values, along with lower RLNE values, indicate greater similarity and better reconstruction quality compared to the fully sampled image.

To comprehensively evaluate the performance of our method, we conducted comparative experiments against several mainstream image reconstruction techniques. These include the traditional PI reconstruction method GRAPPA \cite{GRAPPA}, advanced CS approaches such as pFISTA-SENSE \cite{PFISTA} and AC-LORAKS \cite{ACLORAKS}, as well as implicit neural representation INR-based methods including NERP \cite{NERP} and IMJENSE \cite{IMJENSE}. Here, the CineJENSE is not included in comparison, since
it was proposed for cardiac MRI reconstruction. The degraded version for PI reconstruction of CineJENSE \cite{CINEJENSE}, which is implemented by us, can be regarded as INR-CRISTAL without coil sensitivity map regularization (OURS w.o. Reg.), and we carefully discuss the effect of the sensitivity regularization in Section \ref{sec:reconstruction_constraints} and  \ref{Ablation}.
To ensure a fair and rigorous comparison, we leveraged Bayesian optimization \cite{BAYES} to fine-tune the key parameters of each method. By systematically optimizing their hyperparameters, we maximized the performance of all competing approaches, allowing for a precise and quantitative assessment of their respective strengths and limitations in image reconstruction.
\subsection{Reconstruction with Different Sensitivity Map Constraints}
\label{sec:reconstruction_constraints}

\begin{figure*}
\centering 
\includegraphics[width=\textwidth]{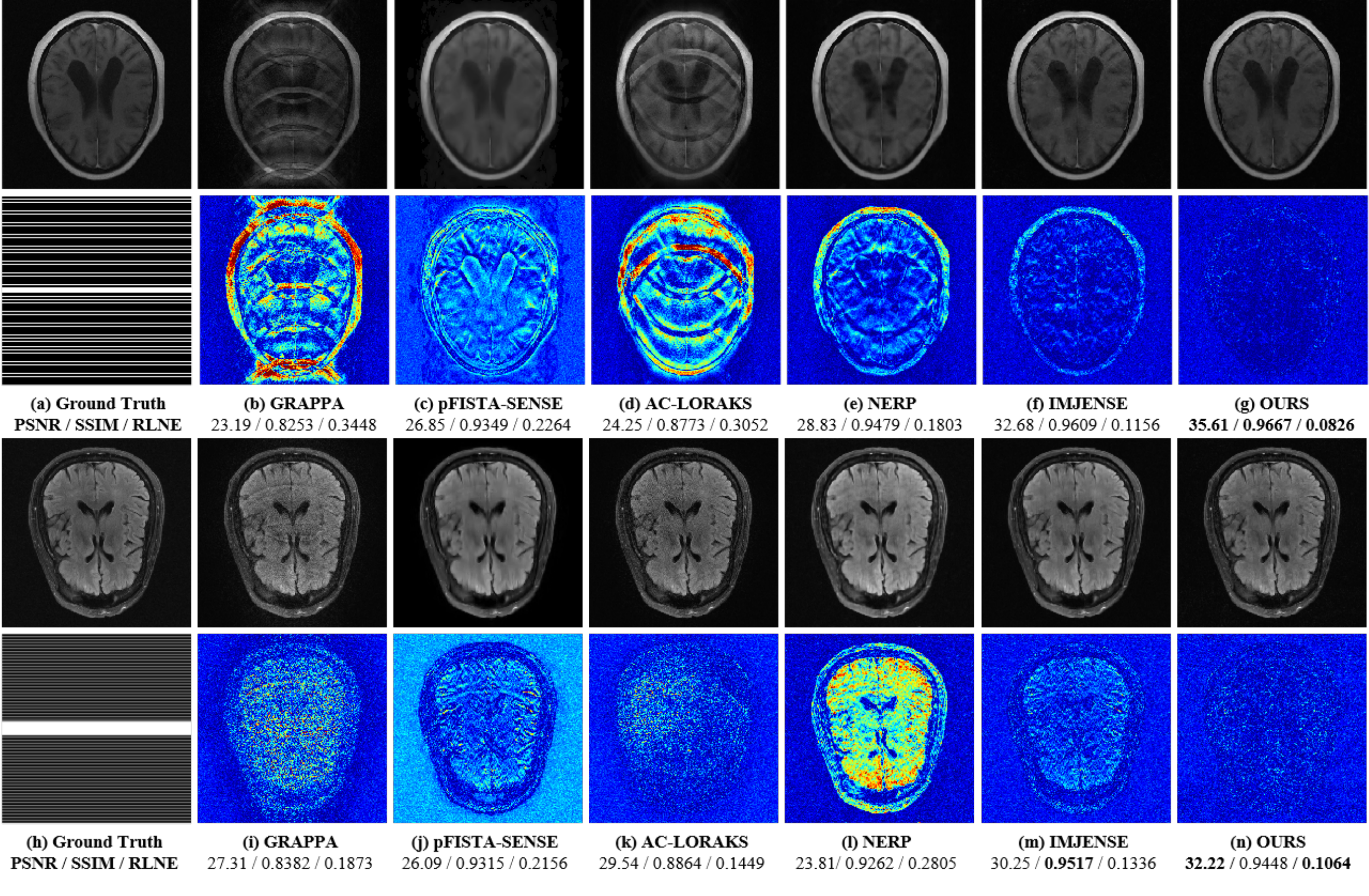}
\caption{Reconstruction results of different methods based on Dataset \Rmnum{1} and Dataset \Rmnum{2}. (a) (or (h)) shows the fully sampled image for Dataset \Rmnum{1} (or Dataset \Rmnum{2}) and the corresponding sampling mask with ACS = 24, R = 5 (ACS = 8). (b)-(g) (or (i)-(n)) show the reconstructed Dataset \Rmnum{1} (or Dataset \Rmnum{2}) images for different methods along with the corresponding error maps (5x). Below each reconstructed image, the evaluation metrics PSNR/ SSIM/ RLNE are listed, with the \textbf{best data} highlighted in bold.}
    \label{fig:main_data_2/3}
\end{figure*}

In this section, we assess the impact of the different regularization
on coil sensitivity map estimation introduced in Section~\ref{Coils Sensitivity constraint}, including regularization with Eq.~\eqref{eq:l1} (OURS-L1), with Eq.~\eqref{eq:lowrank} (OURS-LR), and with Eq.~\eqref{eq:tv} (OURS-TV).

First of all, the proposed method with sensitivity constraints—regardless of the specific type—can reduce the reconstruction error and improve the signal-to-noise ratio compared with the unconstrained one. From Fig.~\ref{fig:l1_lowrank}, it can be observed that the sensitivity maps generated by INR are closer to the ground truth. However, without the smoothness constraint on sensitivity maps, obvious undersampling artifacts remain on the sensitivity maps, and these artifacts further affect the final reconstructed image, causing structural errors and a decrease in PSNR. The methods with sensitivity constraints can effectively resist undersampling artifacts and reduce reconstruction errors.

\begin{figure*}
    \centering 
    \includegraphics[width=\textwidth]{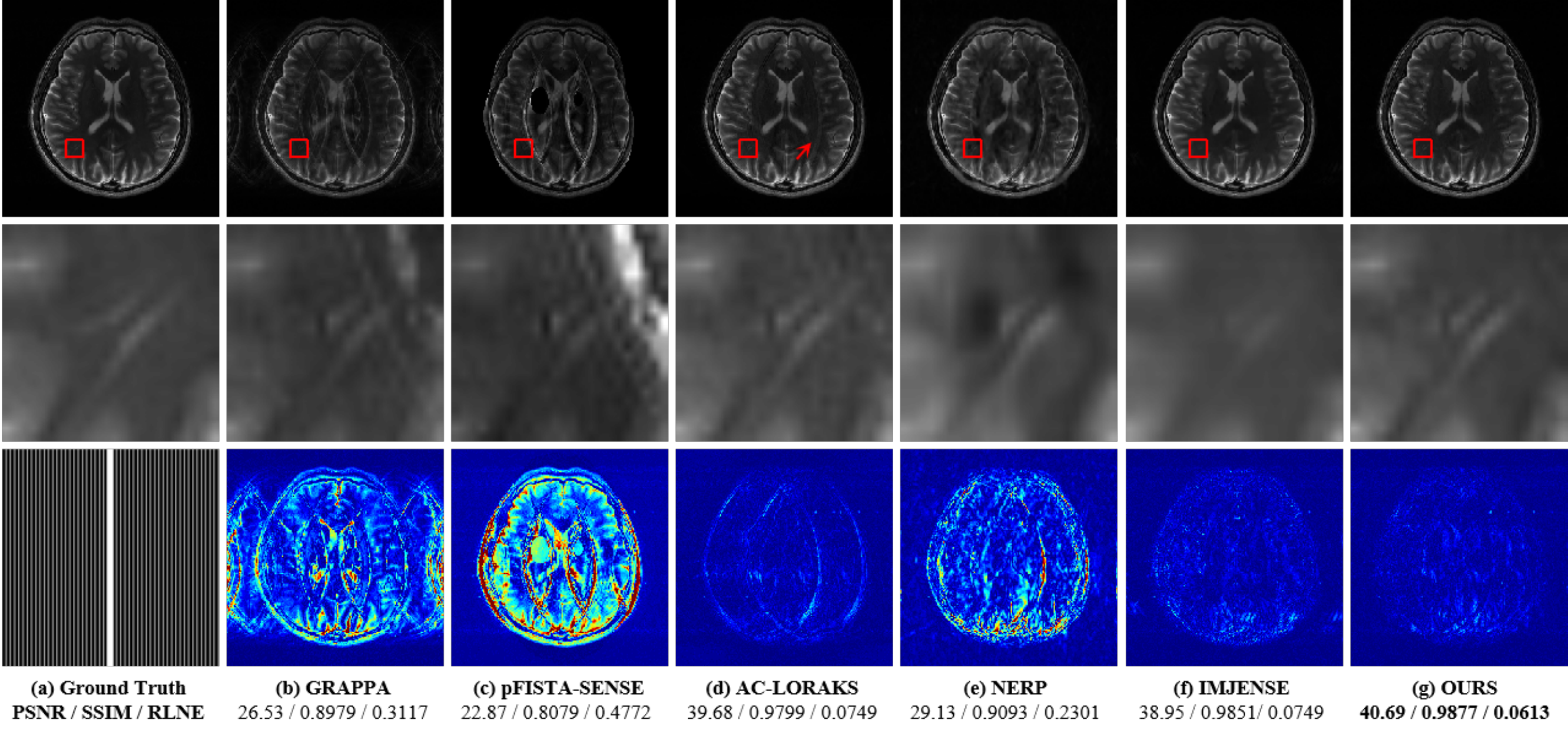}
    \caption{The reconstruction results of different methods based on Dataset \Rmnum{3}. (a) displays the fully sampled image, the zoomed-in view of the red-boxed area, and the sampling mask corresponding to ACS = 8 and R = 5. (b)-(g) show the reconstructed images using different methods, along with the zoomed-in views and the corresponding error maps (5x). The regions marked with red arrows indicate areas with noticeable artifacts. The evaluation metrics PSNR/ SSIM/ RLNE are listed below each reconstructed image, with the \textbf{bolded data} indicating the best performance.}
    \label{fig:main_data_1}
\end{figure*}

\begin{figure*}
\centering 
\includegraphics[width=\textwidth]{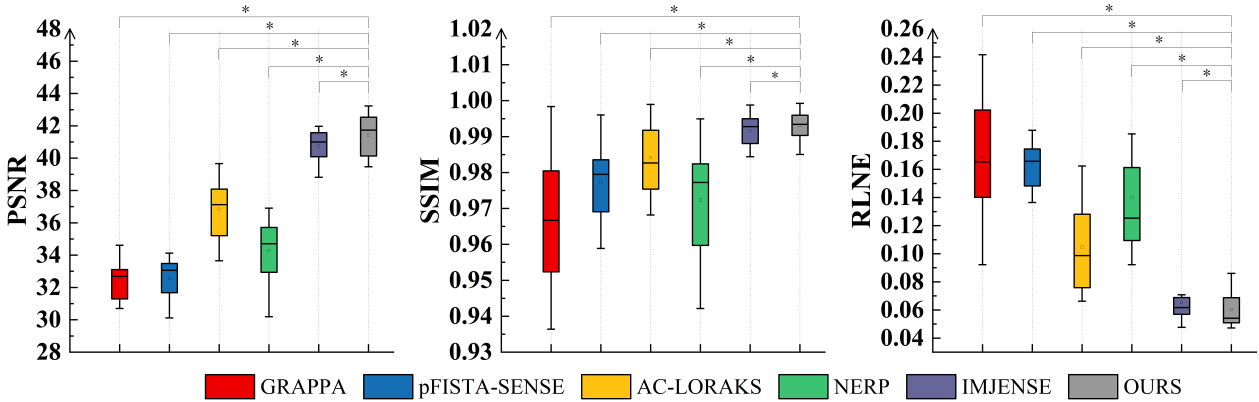}
\caption{Specificity analysis of reconstruction results for Dataset \Rmnum{3} under the ACS = 8 and R = 5 sampling mask. We use the Wilcoxon signed-rank to calculate the p-values of different methods and INR-CRISTAL. When the p-value is less than 0.05, use the symbol $*$ to represent the significant difference between the other methods and INR-CRISTAL.}
    \label{fig:signrank}
\end{figure*}

\begin{figure*}
\centering 
\includegraphics[width=\textwidth]{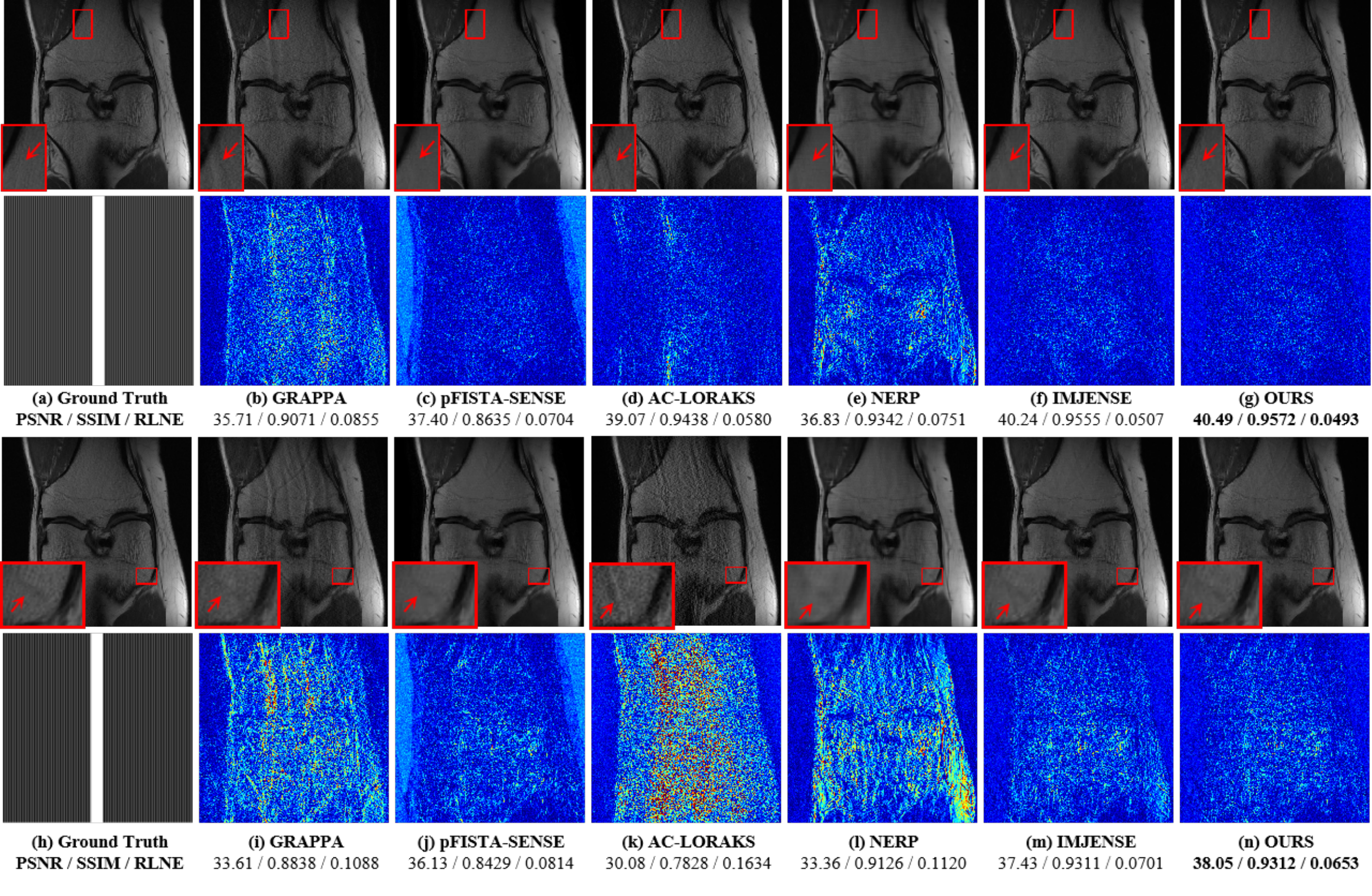}
\caption{Reconstruction results of different methods on the Dataset \Rmnum{4}. (a) (or (h)) shows the fully sampled image and the one-dimensional Cartesian sampling mask at ACS = 24 and R = 4 (or R = 5). (b)-(g) (or (i)-(n)) display the reconstructed images from different methods and their corresponding error maps (10x), with evaluation metrics PSNR, SSIM, and RLNE listed below each reconstruction. The region within the red box shows the corresponding zoomed-in area, and the arrows point to areas with noticeable artifacts. The \textbf{bolded data} indicates the best among the same indicators.}
    \label{fig:main_data_4}
\end{figure*}
The L1-based constraint is essentially to constrain the sparsity of the sensitivity maps on the Fourier basis, however, the Fourier basis and the measurement basis (which is also the Fourier basis) are coherent, which violates the basics of compressed sensing, leading to suboptimal results. The reconstruction metrics on different datasets also illustrate this problem. The reconstruction results of the LR-based constraint are visually cleaner and achieve the highest PSNR value in some cases (Dataset \Rmnum{3}). However, the LR-based regularization method still appears to exhibit undersampling artifacts in some cases (red arrows in Fig.~\ref{fig:l1_lowrank}). In addition, the repeated computation of singular value decompositions for nuclear norm minimization introduces significant computational overhead, resulting in substantially longer reconstruction times compared to the other regularization strategies. In contrast, the TV-based regularization method shows better artifact suppression and structure preservation capabilities, and has achieved results with the highest PSNR and SSIM on multiple datasets.

Considering the computational cost and reconstruction error, the TV-based sensitivity regularization scheme can provide reconstruction with lower error in a shorter reconstruction time. Therefore, we adopt TV-based sensitivity constraints in all subsequent comparative experiments.

\subsection{Reconstruction with different datasets}

In this section, we evaluate the performance of INR-CRISTAL on different datasets. 

In the experiments with Dataset \Rmnum{1} and Dataset \Rmnum{2} (Fig.~\ref{fig:main_data_2/3}), GRAPPA and pFISTA-SENSE fail to reconstruct high-quality images especially when ACS is limited. This is expected, as both methods heavily rely on ACS data to estimate sensitivity maps or interpolation kernels, and insufficient ACS data leads to incomplete calibration. Although NERP and AC-LORAKS manage to recover the overall structure, their reconstructions still exhibit noticeable artifacts. These artifacts arise from challenges in accurately modeling the underlying signal structure under high undersampling conditions. Both IMJENSE and INR-CRISTAL demonstrate superior artifact suppression. However, INR-CRISTAL demonstrates superior reconstruction fidelity by preserving finer image details and yielding lower residual errors, as reflected in its more compact and less noisy residual maps. This advantage is further corroborated by the quantitative metrics, where INR-CRISTAL achieves the highest PSNR and SSIM values and the lowest RLNE, highlighting its ability to balance artifact suppression and structural fidelity effectively.

Moreover, we conducted comparative experiments on Dataset \Rmnum{3}. As shown in Fig.~\ref{fig:main_data_1}, we observed similar trends to those in Dataset \Rmnum{1} and Dataset \Rmnum{2}: both IMJENSE and INR-CRISTAL demonstrate superior artifact suppression. However, as shown in the zoomed-in views, IMJENSE introduces slight blurring in fine structural details, whereas INR-CRISTAL preserves these details more faithfully. INR-CRISTAL effectively suppresses artifacts that other methods fail to remove, maintaining a high level of reconstruction quality. 

To further assess the statistical significance of differences among the methods, we employ the Wilcoxon signed-rank test \cite{Wilcoxon_signed_rankl} (Fig.~\ref{fig:signrank}) on Dataset \Rmnum{3}. The results indicate that when the p-value is below 0.05, the difference between INR-CRISTAL and the other methods is statistically significant, reaffirming INR-CRISTAL’s superior performance.

\begin{table}[htpb]
\caption{Quantitative comparison on  Dataset \Rmnum{4} (37 slices), evaluated by PSNR, SSIM, and RLNE. Values are reported as mean ± standard deviation. \textbf{Bold} values indicate the best performance across methods under each condition.}
\begin{center}
\resizebox{0.48\textwidth}{!}{
\begin{tabular}{
>{\columncolor[HTML]{FFFFFF}}c 
>{\columncolor[HTML]{FFFFFF}}c 
>{\columncolor[HTML]{FFFFFF}}c 
>{\columncolor[HTML]{FFFFFF}}c 
>{\columncolor[HTML]{FFFFFF}}c }
\hline
Dataset \Rmnum{4}     & MASK                                                                                                 & PSNR         & SSIM        & RLNE        \\ \hline
GRAPPA    & \cellcolor[HTML]{FFFFFF}                                                                             & 37.060±1.577 & 0.938±0.032 & 0.083±0.018 \\
pFISTA-SENSE   & \cellcolor[HTML]{FFFFFF}                                                                             & 33.350±1.617 & 0.924±0.036 & 0.130±0.050 \\
AC-LORAKS & \cellcolor[HTML]{FFFFFF}                                                                             & 38.768±1.522 & 0.964±0.016 & 0.069±0.022 \\
NERP      & \cellcolor[HTML]{FFFFFF}                                                                             & 35.545±2.146 & 0.957±0.015 & 0.101±0.029 \\
IMJENSE   & \cellcolor[HTML]{FFFFFF}                                                                             & 39.710±1.096 & 0.974±0.012 & 0.062±0.017 \\
OURS      & \multirow{-6}{*}{\cellcolor[HTML]{FFFFFF}\begin{tabular}[c]{@{}c@{}}R=4\\    \\ ACS=24\end{tabular}}  & \textbf{40.227±1.449} & \textbf{0.977±0.010} & \textbf{0.0559±0.023} \\ \hline
GRAPPA    & \cellcolor[HTML]{FFFFFF}                                                                             & 33.598±2.648 & 0.923±0.042 & 0.126±0.034 \\
pFISTA-SENSE   & \cellcolor[HTML]{FFFFFF}                                                                             & 28.803±3.189 & 0.847±0.091 & 0.226±0.093 \\
AC-LORAKS & \cellcolor[HTML]{FFFFFF}                                                                             & 37.076±2.244 & 0.963±0.018 & 0.086±0.031 \\
NERP      & \cellcolor[HTML]{FFFFFF}                                                                             & 33.646±2.507 & 0.923±0.042 & 0.126±0.034 \\
IMJENSE   & \cellcolor[HTML]{FFFFFF}                                                                             & 38.729±0.962 & 0.971±0.014 & 0.069±0.017 \\
OURS      & \multirow{-6}{*}{\cellcolor[HTML]{FFFFFF}\begin{tabular}[c]{@{}c@{}}R=4\\    \\ ACS=12\end{tabular}}  & \textbf{39.787±1.381} & \textbf{0.976±0.011} & \textbf{0.061±0.018} \\ \hline
GRAPPA    & \cellcolor[HTML]{FFFFFF}                                                                             & 35.346±1.780 & 0.922±0.037 & 0.101±0.017 \\
pFISTA-SENSE   & \cellcolor[HTML]{FFFFFF}                                                                             & 25.864±5.821 & 0.809±0.151 & 0.365±0.244 \\
AC-LORAKS & \cellcolor[HTML]{FFFFFF}                                                                             & 33.067±2.277 & 0.873±0.064 & 0.131±0.022 \\
NERP      & \cellcolor[HTML]{FFFFFF}                                                                             & 33.969±1.631 & 0.939±0.022 & 0.119±0.026 \\
IMJENSE   & \cellcolor[HTML]{FFFFFF}                                                                             & 37.901±1.330 & 0.961±0.019 & 0.076±0.016 \\
OURS      & \multirow{-6}{*}{\cellcolor[HTML]{FFFFFF}\begin{tabular}[c]{@{}c@{}}R=5\\    \\ ACS=24\end{tabular}}  & \textbf{38.738±1.164} & \textbf{0.966±0.016} & \textbf{0.069±0.018} \\ \hline
GRAPPA    & \cellcolor[HTML]{FFFFFF}                                                                             & 33.340±1.883 & 0.914±0.041 & 0.126±0.017 \\
pFISTA-SENSE   & \cellcolor[HTML]{FFFFFF}                                                                             & 22.332±4.859 & 0.747±0.115 & 0.505±0.234 \\
AC-LORAKS & \cellcolor[HTML]{FFFFFF}                                                                             & 32.101±2.272 & 0.867±0.069 & 0.147±0.028 \\
NERP      & \cellcolor[HTML]{FFFFFF}                                                                             & 31.001±1.934 & 0.919±0.030 & 0.167±0.030 \\
IMJENSE   & \cellcolor[HTML]{FFFFFF}                                                                             & 36.591±1.348 & 0.955±0.021 & 0.088±0.016 \\
OURS      & \multirow{-6}{*}{\cellcolor[HTML]{FFFFFF}\begin{tabular}[c]{@{}c@{}}R=5\\    \\ ACS=12\end{tabular}}  & \textbf{37.695±1.120} & \textbf{0.961±0.018} & \textbf{0.077±0.017} \\ \hline
\end{tabular}}
\end{center}
\label{table:KNEE}
\end{table}

As for the knee dataset (Fig.~\ref{fig:main_data_4}), both IMJENSE and INR-CRISTAL effectively suppress artifacts, which remain more pronounced in the GRAPPA and NERP reconstructions. As the acceleration factor increases, noticeable streak artifacts emerge in pFISTA-SENSE and AC-LORAKS, whereas IMJENSE and INR-CRISTAL continue to maintain high reconstruction quality. However, in the zoomed-in regions marked by the red box, IMJENSE introduces subtle details that are not present in the original image, potentially indicating residual artifacts or over-smoothing. In contrast, INR-CRISTAL faithfully preserves structural details without introducing spurious features, further demonstrating its robustness in high-fidelity reconstruction. 

Furthermore, we conducted experiments on Dataset \Rmnum{4} to evaluate the robustness of INR-CRISTAL under different datasets, sampling rates, and ACS configurations. The detailed quantitative results (expressed as mean $±$ standard deviation) are provided in Table \ref{table:KNEE}, further supporting the consistent superiority of INR-CRISTAL in reconstruction performance. In addition, plenty of comparative experiments on different slices of  Datasets \Rmnum{3}, \Rmnum{1}, and \Rmnum{2} (Supplementary Material S1) demonstrated the effectiveness of the proposed method.

\section{Discussions}
\subsection{Robustness of Different ACS and R}
\begin{figure}
\centering 
\includegraphics[width=0.5\textwidth]{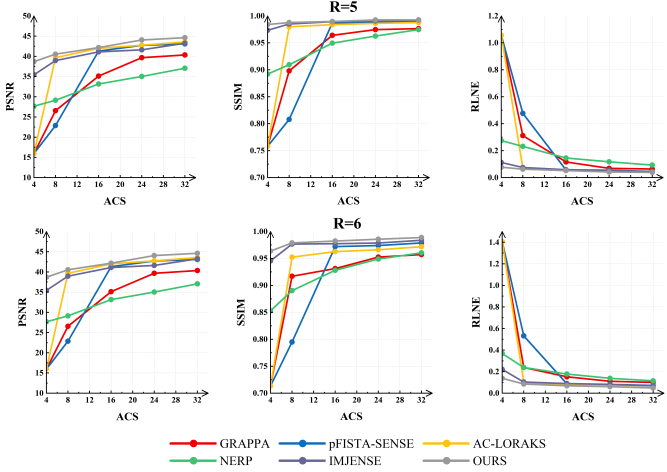}
\caption{Reconstruction results of different methods on Dataset \Rmnum{3}, with ACS as the horizontal axis.}
\label{fig:x_acs}
\end{figure}

In parallel imaging reconstruction, ACS plays a crucial role in ensuring the accuracy of the final image quality \cite{JSENSE}, \cite{JDSI}, \cite{IMJENSE}, \cite{CINEJENSE}. Inaccurate or limited ACS data can negatively impact the estimated sensitivity map, leading to degradation in image reconstruction. We illustrate the variations in PSNR, SSIM, and RLNE as the number of ACS lines changes under acceleration factors R = 5 and R = 6 in Fig.~\ref{fig:x_acs}. Compared to other methods, IMJENSE and INR-CRISTAL demonstrate greater robustness to ACS variations, with INR-CRISTAL consistently achieving the lowest RLNE across all ACS settings, thereby minimizing energy loss.

\begin{figure}
\centering 
\includegraphics[width=0.5\textwidth]{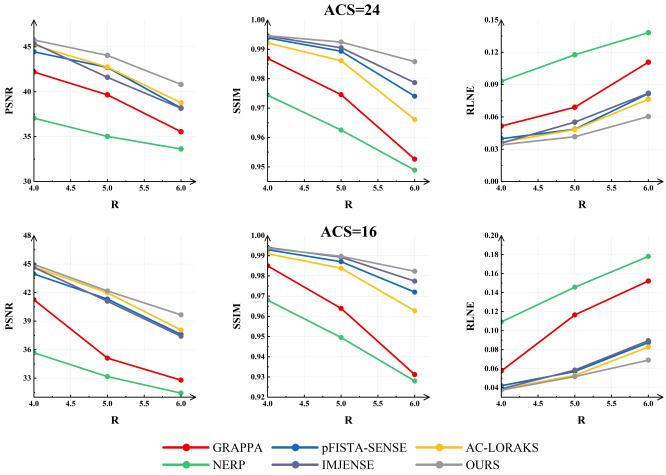}
\caption{Reconstruction results of different methods on Dataset \Rmnum{3}, with R as the horizontal axis.}
\label{fig:x_r}
\end{figure}

In uniform sampling, different acceleration factors R also influence the sampling speed \cite{JDSI}, \cite{IMJENSE}, \cite{CINEJENSE}: the higher the acceleration factor, the lower the sampling rate and the shorter the sampling time. In this study, we aim to maximize R while maintaining reconstruction quality to optimize acquisition efficiency. As R varies under ACS = 24 and ACS = 16, the PSNR, SSIM, and RLNE curves for different methods are presented in Fig.~\ref{fig:x_r}. The results indicate that while all methods perform well at R = 4, INR-CRISTAL consistently minimizes image distortion at R = 6, further validating its stability and superior performance under high acceleration conditions.

\subsection{Reconstruction without ACS}
Obtaining ACS lines requires extended data acquisition time, which increases the risk of motion artifacts from the subject. Consequently, reconstruction methods that do not rely on ACS have garnered significant attention in the research community \cite{PLORAKS}, \cite{calibrationless_SAKE}.

\begin{figure}
\centering 
\includegraphics[width=0.5\textwidth]{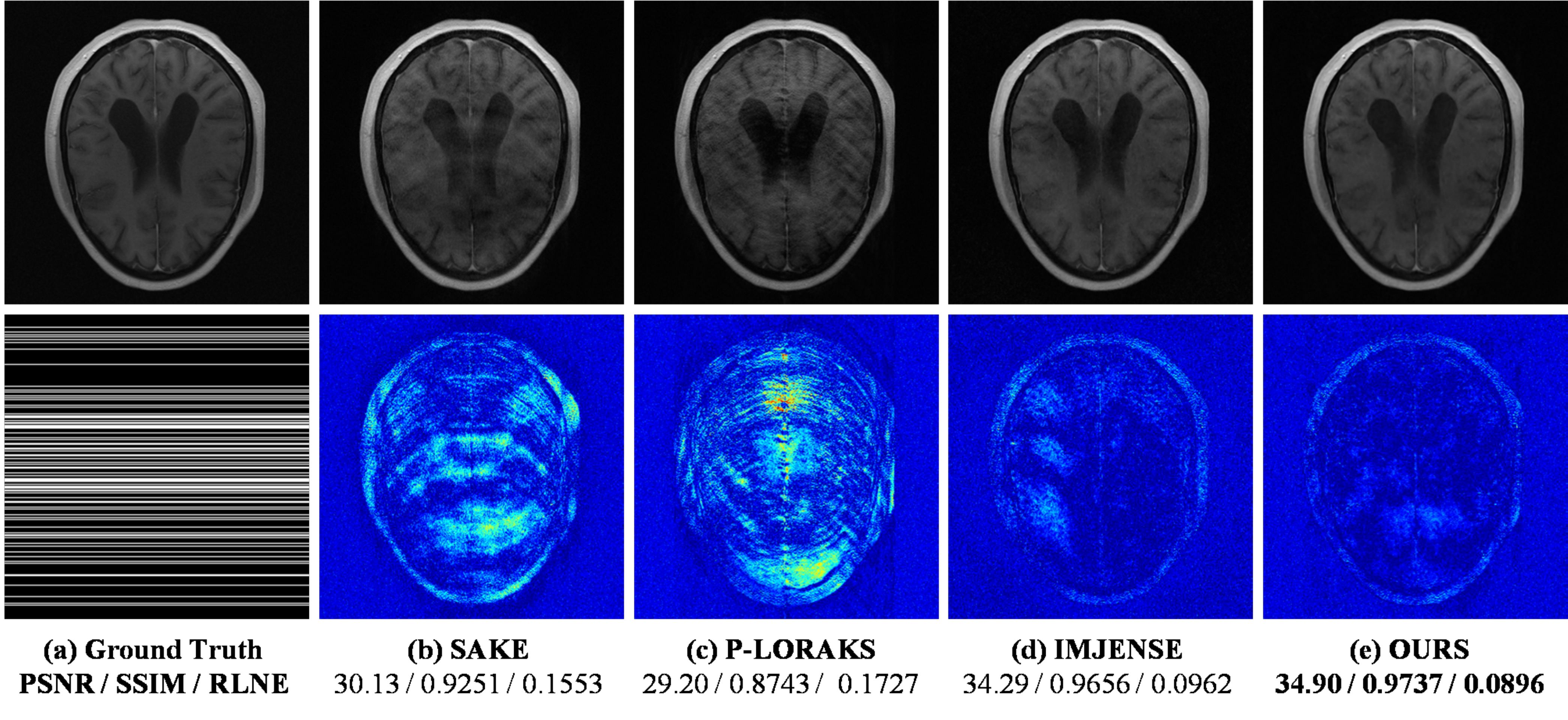}
\caption{Comparison of reconstruction images and error maps (5x) for SAKE, P-LORAKS, IMJENSE and INR-CRISTAL under the non-ACS Gaussian sampling mask with a sampling rate of 0.25 for Dataset \Rmnum{1}. The \textbf{bolded data} indicates the best among the same indicators.}
\label{fig:no_acs}
\end{figure}

To evaluate ACS-free reconstruction performance, we compare the proposed method with P-LORAKS \cite{PLORAKS}, SAKE \cite{calibrationless_SAKE}, and IMJENSE \cite{IMJENSE}. Unlike traditional approaches that estimate the sensitivity map primarily from the central k-space region, INR-CRISTAL leverages continuous functions generated by INR to iteratively refine regularization constraints, enabling effective reconstruction even in the absence of ACS lines. As illustrated in Fig.~\ref{fig:no_acs}, INR-CRISTAL consistently outperforms other ACS-free methods, achieving the lowest RLNE and the highest PSNR across different undersampling acceleration factors. These results further highlight its robustness and adaptability in challenging reconstruction scenarios.

\subsection{Ablation experiment}
\label{Ablation}
To validate the effectiveness of the constraint imposed on the sensitivity map, we conducted ablation experiments on Dataset \Rmnum{3}. As shown in Table \ref{table:6}, when sufficient ACS lines are available, the constraint on the sensitivity map has a relatively limited impact on the reconstruction results. However, it still contributes to an improvement of nearly 0.5 dB in PSNR, which can be attributed to the enhanced consistency of coil sensitivity map estimation, leading to better suppression of aliasing artifacts. In contrast, when the number of ACS lines is limited, the constraint on the sensitivity map plays a crucial role in removing undersampling artifacts from the sensitivity map itself, thereby significantly improving the reconstruction quality. Without this constraint, INR-CRISTAL without sensitivity map regularization struggles to accurately calibrate the sensitivity map during training, as the absence of effective regularization leads to the incorporation of additional artifact information and, consequently, image quality degradation (i.e., overfitting). Overall, INR-CRISTAL achieves the highest PSNR and SSIM values while attaining the lowest RLNE, strongly demonstrating the effectiveness of the adopted sensitivity map constraints.

\begin{table}[htbp]
\caption{Quantitative evaluation metrics of different variants in Dataset \Rmnum{3} ablation experiment.}
\centering
\footnotesize  
\setlength{\tabcolsep}{5pt}  
\renewcommand{\arraystretch}{1.1}  
\begin{tabular}{cccccc}
\toprule
ACS & R & $\mathcal{R}_{TV}$ & PSNR & SSIM & RLNE \\
\midrule
24 & 5 &            & 43.64 & 0.9912 & 0.0436 \\
24 & 5 & \checkmark & \textbf{44.05} & \textbf{0.9925} & \textbf{0.0417} \\
\midrule
24 & 6 &            & 40.46 & 0.9839 & 0.0629 \\
24 & 6 & \checkmark & \textbf{40.81} & \textbf{0.9858} & \textbf{0.0604} \\
\midrule
8 & 5 &             & 38.77 & 0.9842 & 0.0735 \\
8 & 5 & \checkmark  & \textbf{40.53} & \textbf{0.9877} & \textbf{0.0624} \\
\midrule
8 & 6 &             & 35.33 & 0.9695 & 0.1136 \\
8 & 6 & \checkmark  & \textbf{37.83} & \textbf{0.9783} & \textbf{0.0852} \\
\bottomrule
\end{tabular}
\label{table:6}
\end{table}

\section{Conclusion}

In this work, we propose the INR-CRISTAL method for parallel MRI reconstruction, which utilizes the physical characteristics of sensitivity maps by imposing constraints on them to reduce the reconstruction error and remove artifacts when the ACS region is limited. Experimental results demonstrate that the INR-CRISTAL provides accurate sensitivity maps and yields reconstruction with the lowest error. Compared to the state-of-the-art methods, the proposed INR-CRISTAL is superior in terms of artifact removal. Additionally, INR-CRISTAL demonstrates robustness to the ACS region, enabling the lowest reconstruction error with fewer ACS lines and lower sampling rates, thereby boosting its potential for fast imaging applications. Despite its superior performance, INR-CRISTAL still requires a relatively long reconstruction time, and future efforts will focus on accelerating this process while further improving image quality.

\bibliographystyle{IEEEtran}
\IEEEtriggeratref{48}
{\footnotesize
\bibliography{reference}
}

\ifCLASSOPTIONcaptionsoff
  \newpage
\fi

\end{CJK}
\end{document}